\documentclass[english,superscriptaddress,floatfix,twocolumn,prl]{revtex4}
\usepackage[T1]{fontenc}
\usepackage[latin9]{inputenc}
\usepackage{bm}
\usepackage{amsmath}
\usepackage{amssymb}
\usepackage{graphicx}

\makeatletter
\usepackage{comment}

\makeatother

\usepackage{babel}

\newcommand{\prlsec}[1]{\textit{#1.---}}

\begin{document}

\title{A hybrid quantum system formed by trapping atoms in the near-field of a levitated nanosphere}% Force line breaks with \\
\author{A. Hopper}
\author{P. F. Barker}

 \email{p.barker@ucl.ac.uk}
\affiliation{%
 Department of Physics and Astronomy,
 University College London 
}%

\date{\today}% It is always \today, today,

\begin{abstract}
Near-field, radially symmetric optical potentials centred around a levitated nanosphere can be used for sympathetic cooling and for creating a bound nanosphere-atom system analogous to a large molecule. We demonstrate that the long range, Coulomb-like potential produced by a single blue detuned field increases the collisional cross-section by eight orders of magnitude, allowing fast sympathetic cooling of a trapped nanosphere to microKelvin temperatures using cold atoms. By using two optical fields to create a combination of repulsive and attractive potentials, we demonstrate that a cold atom can be bound to a nanosphere creating a new levitated hybrid quantum system suitable for exploring quantum mechanics with massive particles.

%use the optional argument of the \verb+\item+ command to %give the category of each item. 

\end{abstract}

\maketitle

%\tableofcontents

\prlsec{\textbf{Introduction}}
The ability to cool and manipulate the centre-of-mass motion of isolated nanoparticles levitated in vacuum is seen as an important enabling technology for exploring macroscopic quantum mechanics at mass scales well beyond the current state-of-the-art (25$\times$ 10$^3$ amu)\cite{bigmole}. Experiments have been proposed for studying wavefunction collapse, entanglement, decoherence, and for probing the quantum nature of gravity of particles in the $10^6-10^{10}$ amu mass range\cite{oriol1,oriol2,Bateman1,goldwater,Rahman_2019}. As a levitated particle in vacuum is much like a large atomic or molecular system, it can be cooled and trapped using techniques first developed in atomic physics\cite{ritschcavitycooling,rempefeedback}. Over the last ten years, significant progress has been made towards cooling both the centre-of-mass motion and the also the internal temperature of levitated nanoparticles\cite{novotny1,weasel,refrig}. Cavity cooling and feedback cooling to near the motional ground state has very recently been demonstrated, with temperatures reaching the 10 $\mu$K range, bringing these systems well into the quantum regime\cite{groundstate,PhysRevLett.124.013603}. 

Sympathetic cooling of the centre-of-mass motion by elastic collisions with a cold gas is another route towards creating cold trapped nanoparticles.  These methods have been very successful in creating both ultra-cold atoms and molecules that cannot be laser-cooled\cite{Son2020}, with temperatures recently reaching 200 nK. Sympathetic cooling via the coupling of the thermal motion of cold atoms to a levitated nanosphere with a mediating cavity light field has also been proposed\cite{geracicooling} while more recently sympathetic cooling of YIG nanosphere via an ultra-cold spin-polarized gas has been proposed\cite{seberson2019optical}.

Unlike other quantum optomechanical systems, cold levitated nanoparticles can also be released from the trap, allowing matter-wave interferometry and the measurement of small forces, in the absence of perturbing trapping fields.  However, unlike an atomic system, most levitated nanoparticles do not typically have well-defined internal quantum states that can be coherently manipulated and used for interferometry.  A type of Ramsey interferometry using a single nanoscale particle has been proposed. This uses the spin of a single nitrogen vacancy (S=1, N-V$^{-1}$) as a well-controlled quantum system that is embedded within a levitated nanodiamond\cite{bose1}. However, while this general scheme is a promising way to evidence non-classicality in the center-of-mass motion of a massive quantum system, in optical traps it has so far been plagued by technical issues due in large part to the high internal temperature inherent in these levitation schemes\cite{Rahman,delord}. High temperatures lead to phonon broadening of the internal states as well as both motional and spin decoherence, which hamper the use of this system for matter-wave interferometry\cite{entanglementgravity}. Despite these problems, it has recently been used to demonstrate spin cooling\cite{Delord2020}.

\begin{figure}[!t]
\includegraphics[width=1\columnwidth]{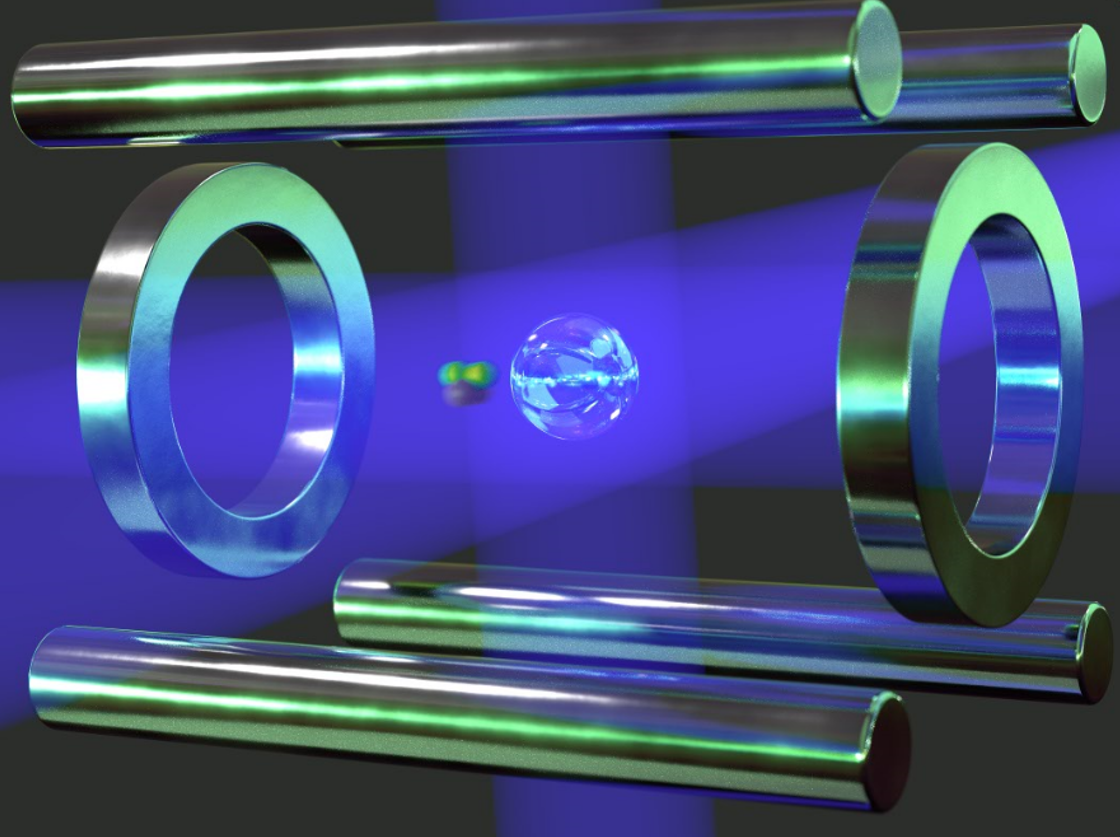}
\caption{A diagram illustrating the creation of a spherically symmetric radial optical potential around a  nanosphere trapped in a Paul trap. The potential, which is always centered on the nanosphere, is created by both the incident and scattered fields from three loosely focused orthogonally polarized optical fields, shown in blue.}
\label{fig:}
\end{figure}

In this letter, we describe a different type of levitated hybrid quantum system, which utilises the good control over both the internal and external degrees of freedom of cold atoms when trapped within optical potentials centered around a nanosphere. The optical potentials are created by both incident and scattered light fields and can be tuned by controlling laser intensity and wavelength.  This system is attractive because atoms can be strongly bound to the nanosphere via the optical fields, but, unlike an NV nanodiamond or quantum dot where the quantum emitter is embedded in the macroscopic object, they are well isolated from the surface and are therefore not subject to phonon broadening or lattice efects. In addition, the optical potentials can be used to increase the collisional cross-section of the nanoparticle enabling rapid thermalization via collisions with cold atoms.
%Polarization forces between charged and neutral particles %are long range and offer the opportunity for enhanced %collisional cooling of nanoparticles with laser cooled %gases. In combination with a repulsive potential created %by the near field scattering of light allows enhance %scattering cross sections and the potential to creation %of massive (10$^9$ amu) molecular ions. Collisions have %been explored widely in the context of buffer gas cooling %of ions and collisions between trapped ions and even %gases within a BEC.  In addition, at low collision %energies it may be possible to create an neutral atom %bound to a nanoparticle creating both a massive and %macroscopic diatomic molecular system. The aim of this %note is to initially explore collisional cooling via an %enhanced collision cross-section due to large %polarization forces.

%This is required to prevent direct contact between an incoming atom and the surface of the nanoparticles which would lead to a release of internal energy of the nanosphere and to motional heating of the trapped nanosphere.%
\prlsec{\textbf{Radially symmetric optical potentials}}
 We consider a dielectric nanosphere that is levitated in a Paul trap in high vacuum\cite{Bullier_2020}. To create a radially symmetric optical potential around the nanosphere it is illuminated with three weakly focused optical fields propagating in orthogonal directions, as shown in figure 1. When the particle is significantly smaller than the wavelength of light, the scattered electric field and intensity can be readily calculated in the Rayleigh approximation\cite{novotnybook}. For a polarizable particle illuminated by an incident optical field $E=E_0 \exp{i(k x-\omega t)}$, propagating in the $x$ direction and polarized along the $z$ direction, the total field due to both the incident and scattered field in the radial and azimuthal directions using spherical polar co-ordinates is given by 
 \begin{equation} \label{eq1}
\begin{split}
 E_r & = C E \cos{\theta} \left[\frac{2}{k^2 r^2}-\frac{2i}{kr}\right]\frac{k^2}{r} + E \cos{\theta}+ c.c\\
E_\theta &  = C E  \sin{\theta} \left[\frac{1}{k^2 r^2}-\frac{ i}{kr}-1\right]\frac{k^2}{r}-E\sin{\theta}+c.c. \\
\end{split}
\end{equation}
For this polarization there is no $\phi$ dependence in the field, and $C=\alpha/4 \pi\epsilon_0 = a^3 (n^2-1)/(n^2+2)$ where $\alpha$ is the polarizability of the nanosphere, $a$ is the radius of the nanoparticle of refractive index $n$, $k=2 \pi/ \lambda$ is the wavevector of the light, and $\lambda$ is its wavelength. 
The total near-field irradiance or intensity around the nanosphere from a single plane wave is then given by

\begin{equation} \label{eq2}
\begin{split}
I_s & = I_0\frac{[4C r^3+r^6+4 C^2(1+k^2 r^2)]}{r^6} \cos^2{\theta} \\
+ &I_0\frac{[r^6- 2 C r^3(1- k^2 r^2)+C^2(1-k^2 r^2+k^4r^4)]}{r^6} \sin^2{\theta}
\\
\end{split}
\end{equation} 
where $I_0=1/2 \epsilon_0 c E_0^2$. Figure 2a shows a plot of the near field intensity around the nanoparticle, with radius $a=40$ nm, refractive index n=1.43 and wavelength 1000 nm. Note that in contrast to the far-field scattered radiation, the near field scattered intensity is maximized along the polarization direction. As this simple analytical solution only applies in the Rayleigh regime when $a<< \lambda$, we confirm that this is a good approximation for this radius by performing numerical simulations by finite difference time domain calculations using the Lumerical software package\cite{lumerical}.  In order to create a spherically symmetric optical near-field potential, we illuminate the particle with three fields, with orthogonal angles of incidence and polarization. When the three optical fields with angles $\theta$ of $cos^{-1}(x/r)$, $cos^{-1}(y/r)$, $cos^{-1}(z/r)$ are added incoherently, the total intensity is given by
\begin{equation} \label{eq1}
\begin{split}
I_T & = I_0\bigg[3+\frac{6C^2}{r^6}+\frac{2 C^2 k^2}{r^4}+\frac{2 C^2k^4}{r^2}+\frac{4 C k^2}{r}\bigg] 
\end{split}
\end{equation}  
We choose the wavevectors of the three incident fields such that $k\approx k_1\approx k_2 \approx k_3$.  A plot of this combined intensity in the y-x plane is shown in figure 1b, which illustrates the radial spherical symmetry around the nanosphere. Note that the radial dependence of the total field is dependent on the wavelength of light through the wavevector $k=\frac{2\pi}{\lambda}$, which is dominated at short range by the $1/r^{6}$ term, and by the $1/r$ term at long range.

% \begin{figure}[h]
%     \centering
%     \includegraphics[width=\columnwidth]{2dslicecombined.png}
%     \caption{An atom within close proximity to a nanosphere within a optical field. The scattering of the light, which is blue detuned with respect to an atomic resonance, creates a near-field optical potential. The total light field is created by two counter propagating fields of orthogonal polarization and detuned by 80 MHz with respect to each.}
%     \label{fig:mesh1}
% \end{figure}

\begin{figure}[!t]
\includegraphics[width=1.2\columnwidth]{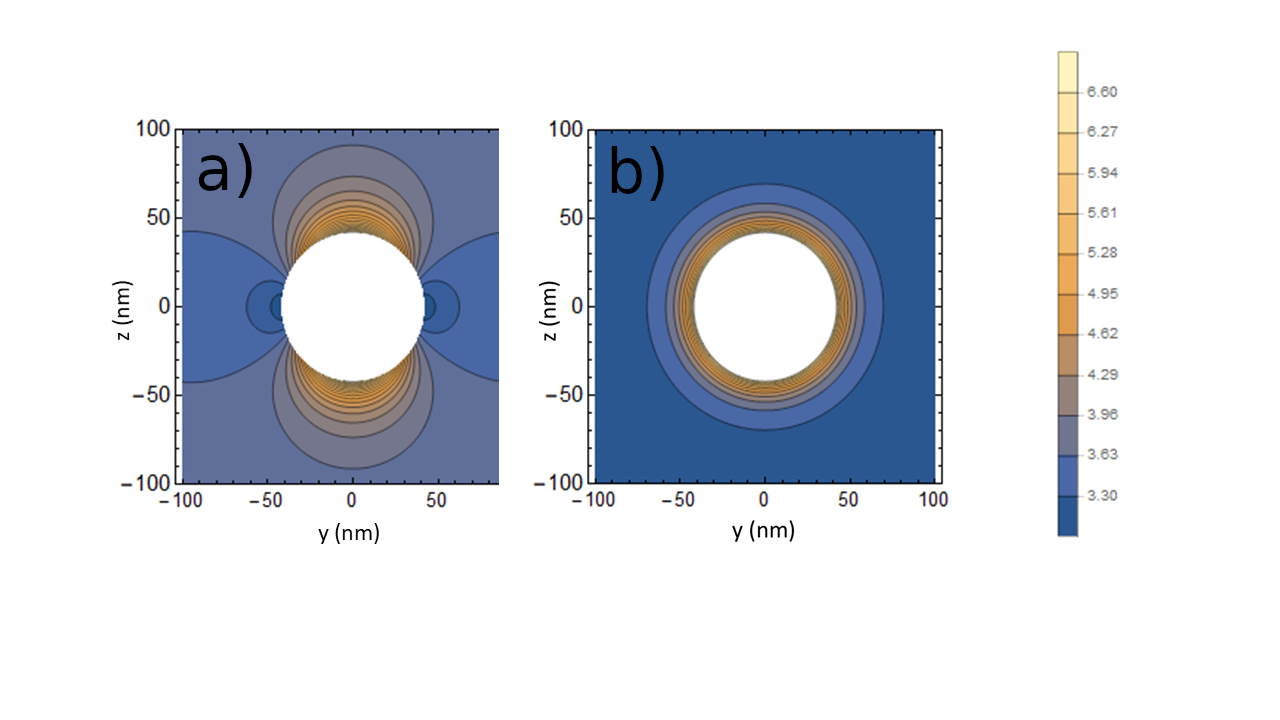}
\vspace*{-20mm}
\caption{The total near-field intensity normalized by the incident beam intensity in the y-z plane. The light of wavelength 1000 nm is incident on a nanosphere of 40 nm radius with and a refractive index of n=1.43. The field inside the nanosphere, shown as the circular white region,is not plotted. a) Is a plot of the intensity when illuminated by a single field propagating in the $x$ direction and polarized in the $z$ direction. b) A plot of the intensity when illuminated by three orthogonally propagating fields with orthogonal polarizations.}
\label{fig:mesh1}
\end{figure}

The interaction of the optical field with either the atom or the nanosphere induces an oscillating dipole moment that leads to an optical dipole potential given by $U_{dip}(r)=-\frac{1}{2}\frac{\alpha I(r)}{\epsilon_0 c}$, where $\alpha$ is the polarizability of an atom or nanosphere\cite{GRIMM200095}. When operating close to an atomic resonance the potential is better described by $U_{dip}(r)=\frac{3 \pi c^2}{2 \omega_0^3}\frac{\Gamma}{\Delta} I(r)$, where $\Delta$ is the detuning of the laser from resonance and $\Gamma$ is the atomic line width. The potential is inversely proportional to the detuning from an atomic resonance, and such potentials are routinely used to trap cold atoms in free space and more recently in the evanescent fields around a tapered optical fiber\cite{atomsfiber1}. As the sign of the potential is determined by its detuning with respect to the atomic resonance or resonances, the detuning can be used to produce both repulsive and attractive potentials. The magnitude of the potential can be tuned by controlling the incident light intensity. The maximum potential at any detuning is eventually limited by the resonant excitation of atoms to other internal states or by increased scattering of light, which can lead to recoil heating. 

To construct a realistic potential, in addition to the applied optical fields, we must also take into account the strong short-range attractive interactions from the Casimir-Polder force and also the charge-induced polarization force between a charged nanosphere and the neutral atom. For charges distributed evenly on the surface, the polarization potential is given by $U=-\frac{1}{(4 \pi \epsilon_0)^2}\frac{\alpha_a q^2}{2r^4}$ where $q$ is the total charge on the nanophere, and $\alpha_a$ is the static polarizability of the atom. For the Casimir-Polder interaction we consider the non-retarded case where the radius of the sphere is significantly less than the resonance wavelengths\cite{cp1}. In this case the potential is given by
\begin{equation} \label{eq1}
\begin{split}
&U_{CP}=- \frac{\hbar}{8 \pi^2 \epsilon_0} \sum_{l=0}^{\infty}(2 l + 1)(l + 1)\\
&\times\frac{a^{2 l + 1}}{r^{2 l + 4}} \int_{0}^{\infty}d \zeta \alpha_a(i \zeta)\frac{\epsilon(i \zeta)-1}{\epsilon(i \zeta)+[(l+1)/l]}
\end{split}
\end{equation} 

where $\zeta$ is the angular frequency, $\alpha_a$ is the atomic polarizability, $\epsilon$ is the permittivity of the nanosphere and $l$ is an integer\cite{cp1}. We calculate numerical solutions based on the properties of the silica and the cold atoms. We find that for less than 100 charges on the nanosphere, which is consistent with Paul trapping experiments\cite{Bullier_2020}, the Casimir-Polder (C-P) force is dominant at all radial distances from the surface of the nanosphere. We therefore only consider the C-P interaction in all subsequent calculations.

Although in principle any laser cooled atom can be trapped around a nanosphere we consider interactions between metastable helium atoms that can be laser cooled in the $2^3S_1$ electronic ground state and a SiO$_2$ nanosphere of radius $a=40$ nm. For the metastable helium atoms we need only consider the few allowed transitions to the $n^3P$ states where $n=2-4$\cite{alphahe}. 

\prlsec{\textbf{Sympathetic cooling}}
We first consider sympathetic cooling of a trapped nanosphere by using a single repulsive optical potential created by three intersecting beams to increase the nanosphere collision cross-section so that the nanoparticle can be sympathetically cooled via collisions with surrounding cold atoms. The repulsive field not only increases the cross-section but also prevents atoms from directly colliding with the nanosphere surface, which will remain at 300 K. These types of collision are important to prevent since they would induce recoil heating of the nanosphere by releasing internal energy into the motion of the nanosphere\cite{heating}. By using a magic wavelength, where the AC stark shift induced by the potential is the same in the upper and lower electronic states, it is possible to simultaneously laser cool the atoms within the optical potential. This will allow continuous cooling of the atoms around the nanosphere. For metastable helium, a repulsive magic wavelength of 318.611 nm has been previously identified\cite{alphahe}.  
   
The full potential for the repulsive optical component $U_R(r)$ and the attractive Casimir-Polder potential is then given by $U_{TR}(r)=U_R(r)+U_{CP}(r)$, where the optical potential is given by the far off-resonance form $U_R(r) = \frac{1}{2} \alpha_a I_T(r)/\epsilon_0 c$, where $\alpha_a=1.33\times 10^{-38}$ Cm$^2$V$^{-1}$ \cite{alphahe} for helium atoms at the magic wavelength.  Figure 3 is a plot of the total potential, as a function of radial distance from the center of the sphere, created by a single beam intensity of $I_0=1\times10^{5}$ W/cm$^2$. This leads to a well depth of 100 K for the silica nanosphere, which is orders of magnitude smaller than Paul trap confining potential\cite{Bullier_2020}. For small collision energies with $E$ in the $E/k_b$=10 $\mu$K range, typical of ultra-cold atomic gases, only the $1/r$ component needs to be considered. 

The scattering of atoms from this potential is then the same form as a repulsive Coulomb potential encountered in classical Rutherford scattering. Without shielding, this type of potential has an infinite total cross-section due to its long range character.  However, as the $1/r$ nature of the optical potential results from an interference between incident and scattered fields, and the incidence field has a finite extent, the 1/r range will be limited to the incident beam size with a spot size of 20 $\mu$m. As we are interested in sympathetic cooling we determine the finite total cross-section, approximating the limited range by using a Yukawa form where $U_Y \approx U_{TR} \exp(-\frac{r}{\mu})$ and where $\mu= 20 \mu$m. This total cross section is plotted in the inset in figure 3 as a function of collision energy in units of $\mu$K. At a collision energy of 10 $\mu$K, this corresponds to a cross-section of $\sigma$ = 6.6$\times$10$^{11}$ nm$^2$, which is more than 8 orders of magnitude larger than the physical cross-section of the 40 nm nanosphere. 
%For this maximum intensity we calculate the average lifetime of atoms in the metastable ground state within %the field from the scattering rate induced by the strongest nearest resonant transition ($2^3S_1 \rightarrow %4^3P_{0,1,2}$). This is calculated to be $\tau=2 Pi/\Gamma_{sc}\approx$100 ms where $\Gamma_{sc}=\frac{3 \pi %c^2}{2\hbar \omega_0^3}\Big(\frac{\Gamma}{\omega_0-\omega}\Big)^2 I_0$.    
\begin{figure}[h]
\centering
\includegraphics[width=0.5\textwidth]{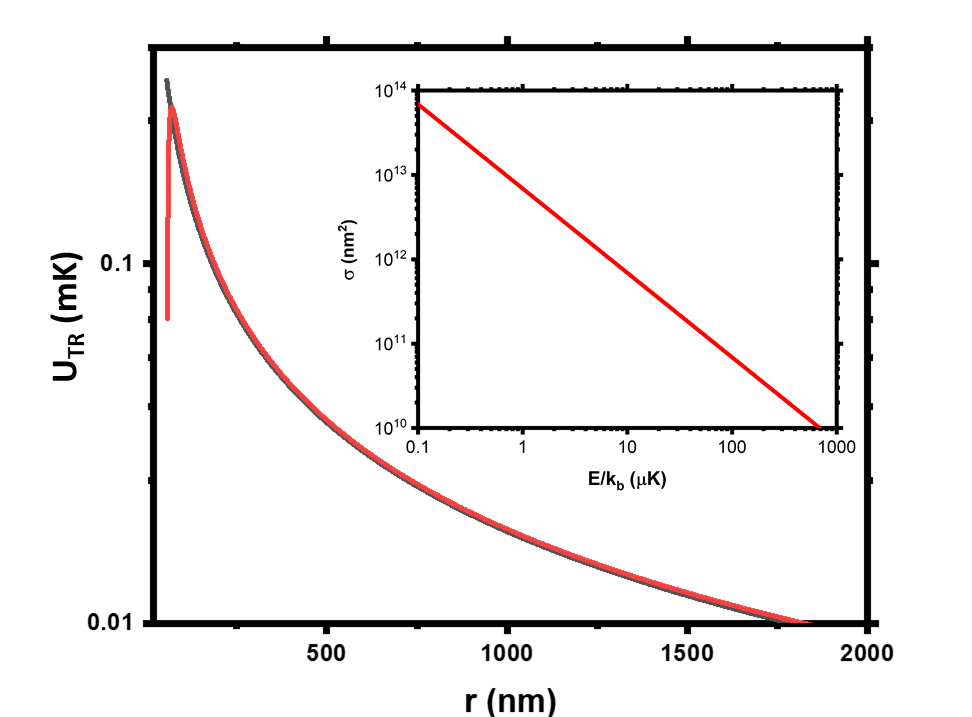}
\caption{The radial potential formed by the combination of an attractive short range potential due to the Casimir-Polder force and a short and long range potential formed by a repulsive optical dipole potential using near field light around a nanoparticle of radius 40 nm. The red line is the full potential and the black line is the Yukawa potential approximation with $\mu$=20  $\mu$m. The inset figure is the classical momentum transfer cross-section for metastable helium atoms when the nanosphere is illuminated by light at the magic wavelength of 318.611 nm.}
\label{fig:mesh1}
\end{figure}

The collisional cooling rate of the nanosphere induced by a bath of cold atoms is approximately given by $\gamma_C=\frac{\xi}{\alpha_c} N <\sigma(v)> v_{th}$, where the brackets denote the speed averaged cross-section, $N$ is the number density of the gas, $v_{th}=\sqrt{\frac{8 k_b}{\pi}(\frac{T_a}{m_a}+\frac{T_n}{m_n})}$ is the thermal speed of the gas, and $\alpha_c$=2.7 is the number of collisions required to thermalize collision partners with equal masses\cite{sympcool}. For unequal masses, as in our case, the cooling rate is reduced by the factor $\xi=\frac{4 m_n m_a}{(m_n+m_a)^2}$, where $m_n = 3.2\times10^8$ amu is the mass of the nanosphere, and $m_a=4.0$ amu is the atomic mass. This large mass difference, which usually makes sympathetic cooling inefficient is compensated for by the large cross-section created by the repulsive optical potential.  For a silica nanosphere of density $\rho$=2000 kg m$^{-3}$ with radius 40 nm and a cold atom gas number density of 5$\times$10$^{12}$ cm$^{-3}$, we determine a cooling/damping rate of $2\pi \times2.3$ kHz at $T_a$=10 $\mu$K, which can be compared with a damping rate of 2$\pi \times $10 $\mu$Hz for the bare sphere. The ultimate temperature that could be reached will be given by additional technical noise sources, but we estimate an effective temperature from $T_e=(\gamma_C T_C+\gamma_H T_H+\gamma_R T_R)/ (\gamma_{C}+\gamma_{H}+\gamma_R)$, where $\gamma_C$ is the the damping rate due to collisions with cold atoms of temperature $T_C$, $\gamma_H$ and $T_H$ are the same quantities due to collision with background atoms, while $\gamma_R$ and $T_R$ are from recoil heating from the light used to form the potential\cite{novotnyrecoil}. We do not include the effect of electrical force noise as this depends on supply, trap geometry and charge on the sphere.  Assuming a background gas pressure of 10$^{-9}$ mbar of air, we derive a temperature of 14 $\mu$K for a cold gas temperature of $T_C$=10 $\mu$K and 2.6$\mu$K  at 1 $\mu$K. Trap frequencies up to at least 50 kHz appear feasible for these small nanoparticles levitated in a Paul trap \cite{doi:10.1021/nn405920k} where, at a temperature of 1 $\mu$K an average phonon occupancy approaching unity could be achieved. In such a scheme ground state cooling may be possible with only small additional feedback cooling. 

Although we have considered sympathetic cooling in a Paul trap we point out that such a scheme could also be used in an optical trap formed by three intersecting beams. Here the intensity is orders of magnitude higher than in the Paul trap and the increased heating rate due to photon recoil would not allow cooling to below one Kelvin.        
%\begin{figure}[h]
%    \centering
%    \includegraphics[width=0.35\textwidth]{allpotentials.png}
%    \caption{The radial potential formed by the combination of an attractive long range component due to %polarization potential from the charged nanoparticle and neutral atom and a short range potential formed by a %repulsive optical dipole potential formed by the near field scattering of the light from the nanoparticle}
%    \label{fig:mesh1}
%\end{figure}

\prlsec{\textbf{Atom-nanosphere bound states}}
We now explore the creation of a bound atom-nanosphere hybrid system using a combination of both a repulsive and an attractive optical potential given by 
\begin{equation} \label{eq1}
\begin{split}
&U_{0}(r)=U_{CP}+3(U_R-U_A)+6\frac{U_R C_R^2-U_AC_A^2}{r^6}\\
&+4\frac{U_R C_R k_R^2-U_A C_A k_A^2}{r}\\
\end{split}
\end{equation} 
where $U_R$ and $U_A$ are the maximum repulsive and attractive well depths in the absence of the scattered fields, and $U_CP$ is due to the Casimir-Polder interaction as before. Note that the $1/r^4$ and the $1/r^2$ terms of equation 3 are not significant and are not included here. Initially ignoring the effect of the CP term, which is only dominant at very short range ($r < 50$ nm), the short range($1/r^{6}$) optical repulsive term constructed from blue detuned light must exceed the corresponding term for attractive red detuned light. In addition, its maximum value must be significantly greater than the energy of the cold atoms to avoid direction collisions with the surface of the sphere. From equation 5 this condition implies that $U_R C_R^2>U_A C_A^2$. However, at long range the attractive ($1/r$) potential created by the red detuned field must be larger than the repulsive component so that $U_R C_R k_R^2<U_A C_A k_A^2$. For a small refractive index variation between the two wavelengths, such that $C_R \approx C_A$, then $k_R^2<k_A^2$ and the wavelength of light used to construct the repulsive optical potential must therefore be larger than that used for the attractive potential ($ \lambda_R^2 > \lambda_A^2$). For a small variation in the refractive index of the nanosphere, we obtain approximate limits of the ratio of the magnitude of the repulsive and attractive potentials given by $1 < \frac{U_R}{U_A} < \frac{\lambda_R^2}{\lambda_A^2}$. Using the $2^3S_1\rightarrow n^3P$ resonances of metastable helium to enhance the optical well depth, we can choose a blue detuned field (5000 GHz) with respect to the $2^3S_1 - 2^3P$ resonance at 1083 nm to create a repulsive potential, and an attractive potential formed by a field that is red detuned by 100 GHz with respect to the $2^3S_1-3^3P$ transition at 389.0 nm.   Our choice of wavelength for both potentials limits the ratio of the repulsive to attractive optical potential in the range $1<U_R/U_A<7.75$. We explored the properties of this potential for creating bound states for $U_R/U_A=2.47$ using well depths of $U_R$=2.7 mK and $U_A$= 14.6 mK. Intensities of approximately $10{^5}$ Wcm$^2$ are required, leading to atom trap lifetimes on the order of 10 ms. 

Figure 4 is plot of the combined potential, $U_0$, which has a well depth of 340 $\mu$ K that is sufficient to trap laser-cooled atoms. We calculate the bound states of this potential and the wavefunctions and allowed energies using the Numerov method\cite{ALLISON1970378}. These are shown in figure 4 for the first four states. This bound potential is similar to the optical potentials created around tapered optical fibers, and the same techniques can in principle be used to load the atoms into this radially symmetric atom-nanosphere potential\cite{atomsfiber1}. Sideband-resolved cooling has been demonstrated around these tapered optical fibers and could also be used to cool the atoms into the ground state of the optical potential around the nanosphere\cite{atomsfiber2}. As the optical potential can be tuned by controlling the relative intensity between the repulsive and  attractive fields scattering, resonances could be used to enhance the cross-section and the binding of cold atoms to the nanosphere. Figure 5 is a calculation of the total cross-section, as a function collision energy using the log-derivative method showing narrow resonances due to partial waves of l=3,7 and higher\cite{JOHNSON1973445}.  The scattering of cold atoms from these potentials and the measurement of energy of these higher energy resonances could be used to characterise the Casimir-Polder potential or to enhance association between the free atom and the nanosphere. Finally, as the trapping volume is small ($\approx$500 nm$^3$) we expect that collisional blockade will result in only one atom captured in the potential, forming a system that is analogous to a massive diatomic molecule. \cite{blockade}.

\begin{figure}[h]
    \centering
  \includegraphics[width=0.45\textwidth]{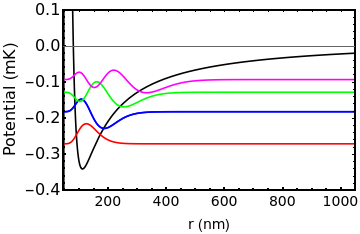}
    \caption{The radial potential formed by the combination of an attractive short-range Casimir-Polder potential and a long range component formed by a repulsive optical dipole potential from the scattering of the light from the nanoparticle. Also shown are the first four wavefunctions and their energies. }
   \label{fig:mesh1}
\end{figure}
\prlsec{\textbf{Conclusion}}
Whilst we have explored interactions with cold metastable helium atoms, the general scheme outlined for both sympathetic cooling and the creation of bound states could be also be carried out with most laser-cooled species. We point out that this scheme only works well for particles where a strong scattered field is created by relatively large nanoscale polarizable particles. Similar fields will be created around large macromolecules of similar size particles, such as a virus which have been trapped in vacuum. Because they are not spherical, as considered here, a more complicated non-spherical potential will be created. However, sympathetic cooling should still be feasible and, as relatively low intensity fields can be used optical damage by absorption of light can be minimised within a Paul trap.  
\begin{figure}[h]
    \centering
  \includegraphics[width=0.45\textwidth]{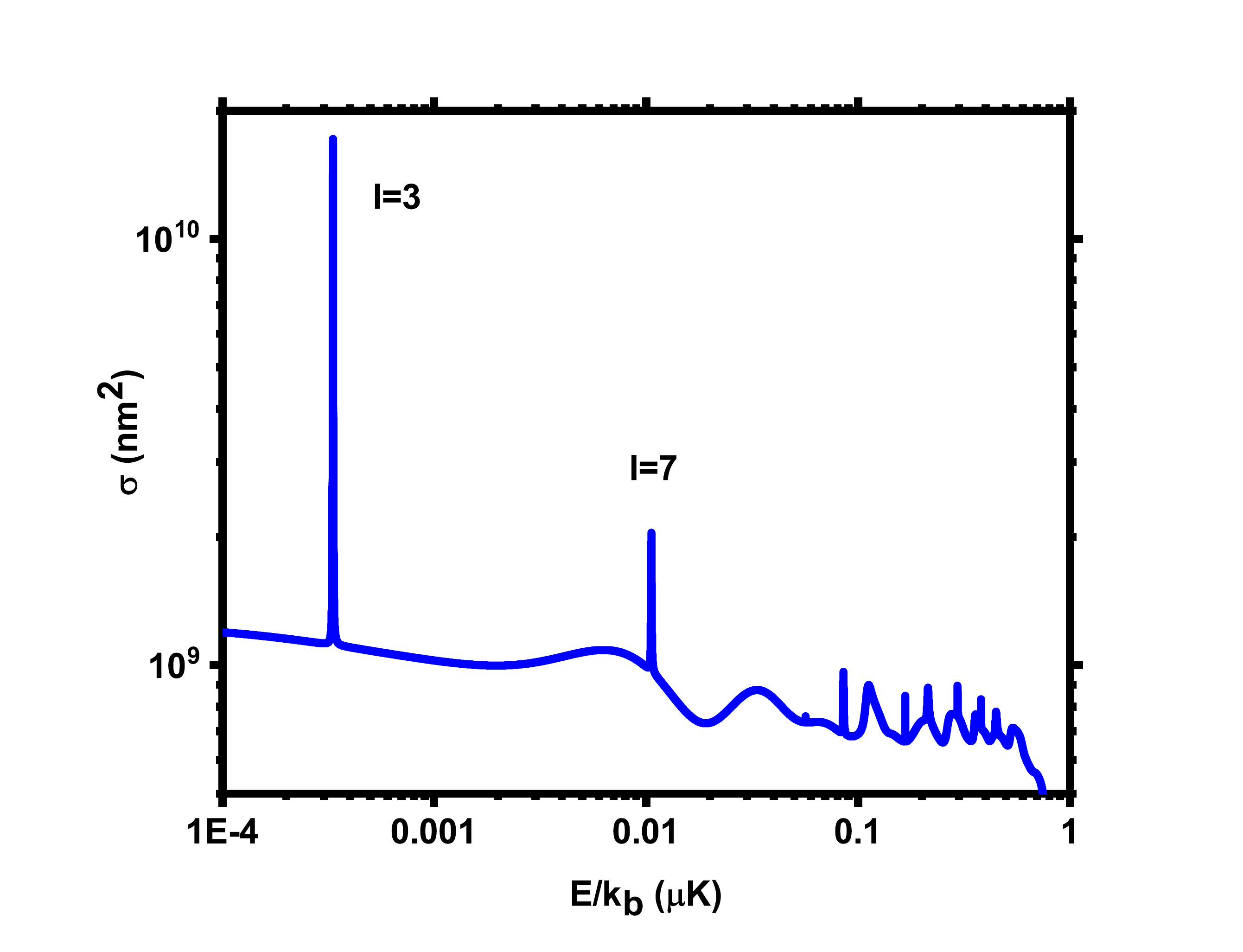}
    \caption{The total scattering cross-section as function of collision energy in units of $\mu$K. Scattering resonances from higher order partial waves could be used to probe the bound state potential and the Casimir-Polder force. Resonances from l=3,7 are indicated.}
   \label{fig:mesh1}
\end{figure}
The hybrid atom-nanosphere system described here may be used as an alternative for the Stern-Gerlach Ramsey type interferometry that has been proposed for NV centres in references\cite{bose1,entanglementgravity}. Here, the metastable helium atom, which is optically bound to the nanosphere, could be placed in a superposition of spin states (m=-1,0,1) by initially preparing the atom in a stretched state (m=-1) via optical pumping before creating a superposition of all spin states\cite{Vassen2016}. By placing the trapped particle in a inhomgeneous magnetic field we can entangle the atomic spin with the motion of the particle via the Zeeman effect converting the spin superposition into a center-of-mass superposition. The creation of macroscopic superpositions of a different optically bound hybrid system will be outlined elsewhere\cite{toroscatstate}.   

%of both red and blue detuned fields we can generate  the 
%the For linear polarized light, and when the particle is %much smaller than the particle size (Rayleigh regime), the %near field is enhanced along the polarization direction of %the light field as shown in figure 1a.  To produce the required radially symmetric intensity gradient, the particle is illuminated by light by three linearly polarized plane wave fields along three orthogonal axes. Figure 1b shows a cross-section in the x-y plane illustrating the isotropic field around the particle.  This is calculated for a 40 nm sphere when illuminated by light at 1000 nm  in the Rayleigh regime. The total radial field intensity from both the incident and scattered fields decays approximately as 1/kR with radial distance from the surface of the sphere.  We point out that as the near field is created by the particle itself, any induced potentials are constant for small displacements in a field with a Gaussian intensity profile.  
 %We also consider the longer-range polarization potential %$U(r)=-\frac{\alpha (N q)^2}{(4 \pi \epsilon_0)^2 2 %r^4}=-\frac{C_4}{2 r^4}$ where $C_4=-\frac{\alpha (N %q)^2}{(4 \pi \epsilon_0)^2}$ is induced when charges are %present on the sphere. This potential can be ignored for a %few charges but must be taken into account for x or more %charges.

%Equation x shows that the short range $r^-6$ has no %wavelength dependence and is only dependent on field %intensity while the magnitude of the longer range potentials %($r^{-n}$, $n<6$) depends on both the intensity and %wavelength.    

 % 
\prlsec{\textbf{Acknowledgements}}The authors would like to acknowledge useful discussions with M. Toro\v{s}, S. Bose, A. Pontin and A. Rahman. The authors acknowledge funding from the EPSRC Grant No. EP/N031105/1 and the H2020-EU.1.2.1 TEQ project Grant agreement ID: 766900.


\begin{thebibliography}{10}
	\providecommand{\url}[1]{\texttt{#1}}
	\providecommand{\urlprefix}{}
	\providecommand{\arxivId}[2][]{\url{#2}}
	
	\bibitem{bigmole}
	Y.~Y. Fein, P.~Geyer, P.~Zwick \emph{et~al.}
	\newblock \href{https://doi.org/10.1038/s41567-019-0663-9}{Quantum
		superposition of molecules beyond 25 kda}.
	\newblock \emph{Nat. Phys.}, \textbf{15}, 1242 (2019).
	
	\bibitem{oriol1}
	O.~Romero-Isart, M.~L. Juan, R.~Quidant \emph{et~al.}
	\newblock \href{https://doi.org/10.1088%2F1367-2630%2F12%2F3%2F033015}{Toward
		quantum superposition of living organisms}.
	\newblock \emph{New Journal of Physics}, \textbf{12}, 033015 (2010).
	
	\bibitem{oriol2}
	O.~Romero-Isart.
	\newblock \href{https://link.aps.org/doi/10.1103/PhysRevA.84.052121}{Quantum
		superposition of massive objects and collapse models}.
	\newblock \emph{Phys. Rev. A}, \textbf{84}, 052121 (2011).
	
	\bibitem{Bateman1}
	J.~Bateman, S.~Nimmrichter, K.~Hornberger \emph{et~al.}
	\newblock \href{https://doi.org/10.1038/ncomms5788}{Near-field interferometry
		of a free-falling nanoparticle from a point-like source}.
	\newblock \emph{Nature Communications}, \textbf{5} (2014).
	
	\bibitem{goldwater}
	D.~Goldwater, M.~Paternostro, and P.~F. Barker.
	\newblock \href{https://link.aps.org/doi/10.1103/PhysRevA.94.010104}{Testing
		wave-function-collapse models using parametric heating of a trapped
		nanosphere}.
	\newblock \emph{Phys. Rev. A}, \textbf{94}, 010104 (2016).
	
	\bibitem{Rahman_2019}
	A.~T. M.~A. Rahman.
	\newblock \href{https://doi.org/10.1088%2F1367-2630%2Fab4fb3}{Large spatial
		schrödinger cat state using a levitated ferrimagnetic nanoparticle}.
	\newblock \emph{New Journal of Physics}, \textbf{21}, 113011 (2019).
	
	\bibitem{ritschcavitycooling}
	P.~Horak, G.~Hechenblaikner, K.~M. Gheri \emph{et~al.}
	\newblock
	\href{https://link.aps.org/doi/10.1103/PhysRevLett.79.4974}{Cavity-induced
		atom cooling in the strong coupling regime}.
	\newblock \emph{Phys. Rev. Lett.}, \textbf{79}, 4974 (1997).
	
	\bibitem{rempefeedback}
	M.~Koch, C.~Sames, A.~Kubanek \emph{et~al.}
	\newblock
	\href{https://link.aps.org/doi/10.1103/PhysRevLett.105.173003}{Feedback
		cooling of a single neutral atom}.
	\newblock \emph{Phys. Rev. Lett.}, \textbf{105}, 173003 (2010).
	
	\bibitem{novotny1}
	F.~Tebbenjohanns, M.~Frimmer, A.~Militaru \emph{et~al.}
	\newblock \href{https://link.aps.org/doi/10.1103/PhysRevLett.122.223601}{Cold
		damping of an optically levitated nanoparticle to microkelvin temperatures}.
	\newblock \emph{Phys. Rev. Lett.}, \textbf{122}, 223601 (2019).
	
	\bibitem{weasel}
	N.~Kiesel, F.~Blaser, U.~Deli{\'c} \emph{et~al.}
	\newblock \href{https://www.pnas.org/content/110/35/14180}{Cavity cooling of an
		optically levitated submicron particle}.
	\newblock \emph{Proceedings of the National Academy of Sciences}, \textbf{110},
	14180 (2013).
	
	\bibitem{refrig}
	A.~T. M.~A. Rahman and P.~F. Barker.
	\newblock \href{https://doi.org/10.1038/s41566-017-0005-3}{Laser refrigeration,
		alignment and rotation of levitated yb3+:ylf nanocrystals}.
	\newblock \emph{Nature Photonics}, \textbf{11}, 634 (2017).
	
	\bibitem{groundstate}
	U.~Deli{\'c}, M.~Reisenbauer, K.~Dare \emph{et~al.}
	\newblock \href{https://science.sciencemag.org/content/367/6480/892}{Cooling of
		a levitated nanoparticle to the motional quantum ground state}.
	\newblock \emph{Science}, \textbf{367}, 892 (2020).
	
	\bibitem{PhysRevLett.124.013603}
	F.~Tebbenjohanns, M.~Frimmer, V.~Jain \emph{et~al.}
	\newblock
	\href{https://link.aps.org/doi/10.1103/PhysRevLett.124.013603}{Motional
		sideband asymmetry of a nanoparticle optically levitated in free space}.
	\newblock \emph{Phys. Rev. Lett.}, \textbf{124}, 013603 (2020).
	
	\bibitem{Son2020}
	H.~Son, J.~J. Park, W.~Ketterle \emph{et~al.}
	\newblock \href{https://doi.org/10.1038/s41586-020-2141-z}{Collisional cooling
		of ultracold molecules}.
	\newblock \emph{Nature}, \textbf{580}, 197 (2020).
	
	\bibitem{geracicooling}
	G.~Ranjit, C.~Montoya, and A.~A. Geraci.
	\newblock \href{https://link.aps.org/doi/10.1103/PhysRevA.91.013416}{Cold atoms
		as a coolant for levitated optomechanical systems}.
	\newblock \emph{Phys. Rev. A}, \textbf{91}, 013416 (2015).
	
	\bibitem{seberson2019optical}
	T.~{Seberson}, P.~{Ju}, J.~{Ahn} \emph{et~al.}
	\newblock {Optical levitation of a YIG nanoparticle and simulation of
		sympathetic cooling via coupling to a cold atomic gas}.
	\newblock \emph{arXiv e-prints}, arXiv:1910.05371 (2019).
	
	\bibitem{bose1}
	M.~Scala, M.~S. Kim, G.~W. Morley \emph{et~al.}
	\newblock
	\href{https://link.aps.org/doi/10.1103/PhysRevLett.111.180403}{Matter-wave
		interferometry of a levitated thermal nano-oscillator induced and probed by a
		spin}.
	\newblock \emph{Phys. Rev. Lett.}, \textbf{111}, 180403 (2013).
	
	\bibitem{Rahman}
	A.~T. M.~A. Rahman, A.~C. Frangeskou, M.~S. Kim \emph{et~al.}
	\newblock \href{https://doi.org/10.1038/srep21633}{Burning and graphitization
		of optically levitated nanodiamonds in vacuum}.
	\newblock \emph{Scientific Reports}, \textbf{6}, 21633 (2016).
	
	\bibitem{delord}
	T.~Delord, L.~Nicolas, M.~Bodini \emph{et~al.}
	\newblock \href{https://doi.org/10.1063/1.4991670}{Diamonds levitating in a
		paul trap under vacuum: Measurements of laser-induced heating via nv center
		thermometry}.
	\newblock \emph{Applied Physics Letters}, \textbf{111}, 013101 (2017).
	
	\bibitem{entanglementgravity}
	S.~Bose, A.~Mazumdar, G.~W. Morley \emph{et~al.}
	\newblock \href{https://link.aps.org/doi/10.1103/PhysRevLett.119.240401}{Spin
		entanglement witness for quantum gravity}.
	\newblock \emph{Phys. Rev. Lett.}, \textbf{119}, 240401 (2017).
	
	\bibitem{Delord2020}
	T.~Delord, P.~Huillery, L.~Nicolas \emph{et~al.}
	\newblock \href{https://doi.org/10.1038/s41586-020-2133-z}{Spin-cooling of the
		motion of a trapped diamond}.
	\newblock \emph{Nature}, \textbf{580}, 56 (2020).
	
	\bibitem{Bullier_2020}
	N.~P. Bullier, A.~Pontin, and P.~F. Barker.
	\newblock \href{https://doi.org/10.1088%2F1361-6463%2Fab71a7}{Characterisation
		of a charged particle levitated nano-oscillator}.
	\newblock \emph{Journal of Physics D: Applied Physics}, \textbf{53}, 175302
	(2020).
	
	\bibitem{novotnybook}
	L.~Novotny and B.~Hecht.
	\newblock \href{http://dx.doi.org/10.1017/CBO9780511813535}{\emph{Principles of
			Nano-Optics}}.
	\newblock Cambridge University Press (2006).
	
	\bibitem{lumerical}
	Lumerical inc. fdtd 3d electromagnetic simulator.
	\newblock \url{ https://www.lumerical.com/products/} (2019).
	
	\bibitem{GRIMM200095}
	R.~Grimm, M.~Weidemüller, and Y.~B. Ovchinnikov.
	\newblock
	\href{http://www.sciencedirect.com/science/article/pii/S1049250X0860186X}{\emph{Optical
			Dipole Traps for Neutral Atoms}}, volume~42 of \emph{Advances In Atomic,
		Molecular, and Optical Physics}.
	\newblock Academic Press (2000).
	
	\bibitem{atomsfiber1}
	E.~Vetsch, D.~Reitz, G.~Sagu\'e \emph{et~al.}
	\newblock
	\href{https://link.aps.org/doi/10.1103/PhysRevLett.104.203603}{Optical
		interface created by laser-cooled atoms trapped in the evanescent field
		surrounding an optical nanofiber}.
	\newblock \emph{Phys. Rev. Lett.}, \textbf{104}, 203603 (2010).
	
	\bibitem{cp1}
	J.~L. Hemmerich, R.~Bennett, T.~Reisinger \emph{et~al.}
	\newblock \href{https://link.aps.org/doi/10.1103/PhysRevA.94.023621}{Impact of
		casimir-polder interaction on poisson-spot diffraction at a dielectric
		sphere}.
	\newblock \emph{Phys. Rev. A}, \textbf{94}, 023621 (2016).
	
	\bibitem{alphahe}
	R.~P. M. J.~W. Notermans, R.~J. Rengelink, K.~A.~H. van Leeuwen \emph{et~al.}
	\newblock \href{https://link.aps.org/doi/10.1103/PhysRevA.90.052508}{Magic
		wavelengths for the
		$2{\phantom{\rule{4pt}{0ex}}}^{3}s\ensuremath{\rightarrow}2{\phantom{\rule{4pt}{0ex}}}^{1}s$
		transition in helium}.
	\newblock \emph{Phys. Rev. A}, \textbf{90}, 052508 (2014).
	
	\bibitem{heating}
	J.~Millen, T.~Deesuwan, P.~Barker \emph{et~al.}
	\newblock \href{https://doi.org/10.1038/nnano.2014.82}{Nanoscale temperature
		measurements using non-equilibrium brownian dynamics of a levitated
		nanosphere}.
	\newblock \emph{Nature Nanotechnology}, \textbf{11}, 425 (2014).
	
	\bibitem{sympcool}
	V.~V. Ivanov, A.~Khramov, A.~H. Hansen \emph{et~al.}
	\newblock
	\href{https://link.aps.org/doi/10.1103/PhysRevLett.106.153201}{Sympathetic
		cooling in an optically trapped mixture of alkali and spin-singlet atoms}.
	\newblock \emph{Phys. Rev. Lett.}, \textbf{106}, 153201 (2011).
	
	\bibitem{novotnyrecoil}
	V.~Jain, J.~Gieseler, C.~Moritz \emph{et~al.}
	\newblock \href{https://link.aps.org/doi/10.1103/PhysRevLett.116.243601}{Direct
		measurement of photon recoil from a levitated nanoparticle}.
	\newblock \emph{Phys. Rev. Lett.}, \textbf{116}, 243601 (2016).
	
	\bibitem{doi:10.1021/nn405920k}
	D.~M. Bell, C.~R. Howder, R.~C. Johnson \emph{et~al.}
	\newblock \href{https://doi.org/10.1021/nn405920k}{Single cdse/zns nanocrystals
		in an ion trap: Charge and mass determination and photophysics evolution with
		changing mass, charge, and temperature}.
	\newblock \emph{ACS Nano}, \textbf{8}, 2387 (2014).
	\newblock PMID: 24410129.
	
	\bibitem{ALLISON1970378}
	A.~Allison.
	\newblock
	\href{http://www.sciencedirect.com/science/article/pii/0021999170900379}{The
		numerical solution of coupled differential equations arising from the
		schrödinger equation}.
	\newblock \emph{Journal of Computational Physics}, \textbf{6}, 378  (1970).
	
	\bibitem{atomsfiber2}
	Y.~Meng, A.~Dareau, P.~Schneeweiss \emph{et~al.}
	\newblock
	\href{https://link.aps.org/doi/10.1103/PhysRevX.8.031054}{Near-ground-state
		cooling of atoms optically trapped 300 nm away from a hot surface}.
	\newblock \emph{Phys. Rev. X}, \textbf{8}, 031054 (2018).
	
	\bibitem{JOHNSON1973445}
	B.~Johnson.
	\newblock
	\href{http://www.sciencedirect.com/science/article/pii/0021999173900491}{The
		multichannel log-derivative method for scattering calculations}.
	\newblock \emph{Journal of Computational Physics}, \textbf{13}, 445  (1973).
	
	\bibitem{blockade}
	N.~Schlosser, G.~Reymond, and P.~Grangier.
	\newblock
	\href{https://link.aps.org/doi/10.1103/PhysRevLett.89.023005}{Collisional
		blockade in microscopic optical dipole traps}.
	\newblock \emph{Phys. Rev. Lett.}, \textbf{89}, 023005 (2002).
	
	\bibitem{Vassen2016}
	W.~Vassen, R.~P. M. J.~W. Notermans, R.~J. Rengelink \emph{et~al.}
	\newblock \href{https://doi.org/10.1007/s00340-016-6563-0}{Ultracold metastable
		helium: Ramsey fringes and atom interferometry}.
	\newblock \emph{Applied Physics B}, \textbf{122}, 289 (2016).
	
	\bibitem{toroscatstate}
	M.~Toros, S.~Bose, and P.~F. Barker.
	\newblock Creation and evidencing of atom-nanoparticle schr\"{o}dinger cats, in
	preparation (2019).
	
\end{thebibliography}
\end{document}